\documentclass{article}
\usepackage[preprint]{neurips_2024}

\usepackage{fancyhdr} 
\usepackage{blindtext} 
\usepackage{makecell}
\usepackage{natbib}

\usepackage{fontawesome5} 

\usepackage[utf8]{inputenc} 
\usepackage[T1]{fontenc}    
\usepackage{hyperref}       
\usepackage{url}            
\usepackage{booktabs}       
\usepackage{amsfonts}       
\usepackage{nicefrac}       
\usepackage{wrapfig}
\usepackage{xcolor}         
\usepackage{amsmath}
\usepackage{float}
\usepackage{tabularray}
\usepackage{multirow}
\usepackage{booktabs, caption, array}
\usepackage{amssymb}
\usepackage{lipsum}         
\usepackage{CJKutf8}        
\usepackage{nomencl}
\usepackage{glossaries}
\usepackage{xspace}
\usepackage{multirow}
\usepackage{graphicx}
\usepackage{bbding}
\usepackage{pifont}
\usepackage{algorithm}
\usepackage{enumitem}
\usepackage{multicol}
\usepackage{caption}
\usepackage{array}
\usepackage{fp}
\usepackage{multirow}   
\usepackage{makecell}
\usepackage{flushend}
\usepackage{cases}
\usepackage{color}
\usepackage{algpseudocode}
\usepackage{listings}
\usepackage{mathrsfs}
\usepackage{longtable}
\usepackage{colortbl}
\usepackage{bm}
\usepackage{caption}
\usepackage{natbib}
\usepackage{tabularx}

\usepackage{subcaption} 
\setlist[itemize]{noitemsep, topsep=0pt}
\usepackage{arydshln}
\usepackage{lipsum}
\definecolor{codegreen}{rgb}{0,0.3,0.6}
\definecolor{codegray}{rgb}{0.5,0.5,0.5}

\newcommand{\ie}{\emph{i.e.,}\xspace}
\newcommand{\eg}{\emph{e.g.,}\xspace}

\newcommand{\paratitle}[1]{\vspace{1.5ex}\noindent\textbf{#1}}
\newcommand{\wrt}{w.r.t.\xspace}

\newcommand{\ignore}[1]{}

\usepackage[most]{tcolorbox} 
\usepackage{microtype}      
\definecolor{darkorange}{RGB}{255, 140, 0}
\definecolor{lightgreen}{RGB}{145, 204, 117}
\definecolor{lightyellow}{RGB}{250, 200, 88}
\definecolor{lightred}{RGB}{238, 102, 102}
\definecolor{lightblue}{RGB}{115, 192, 222}
\newtcolorbox{promptbox}[3][Judge Prompt]{
colback=black!5!white,
arc=5pt, 
boxrule=0.5pt,
fonttitle=\bfseries,
title=#1, 
before upper={\small}, fontupper=\fontfamily{ptm}\selectfont,
colframe=#2,
label=#3,
}
\definecolor{gray_1}{HTML}{B7B7B7}
\definecolor{gray_2}{HTML}{F0F0F0} 
\definecolor{frame_blue}{HTML}{A9D18E}

\newtcolorbox[auto counter, number within=section]{PromptBoxNew}[2][]{
    enhanced,
    breakable,
    colback=gray_2, 
    colframe=gray_1,
    coltitle=white,
    fontupper=\small,
    fonttitle=\bfseries,
    title={#2}, 
    label={#1},
    arc=2pt,
    boxrule=1pt,
    left=2mm, right=2mm, top=2mm, bottom=2mm,
}

\usepackage{etoolbox}
\makeatletter
\patchcmd{\@maketitle}
  {\@toptitlebar}
  {%
    \hbox to \textwidth{%
    
    \includegraphics[height=0.75cm]{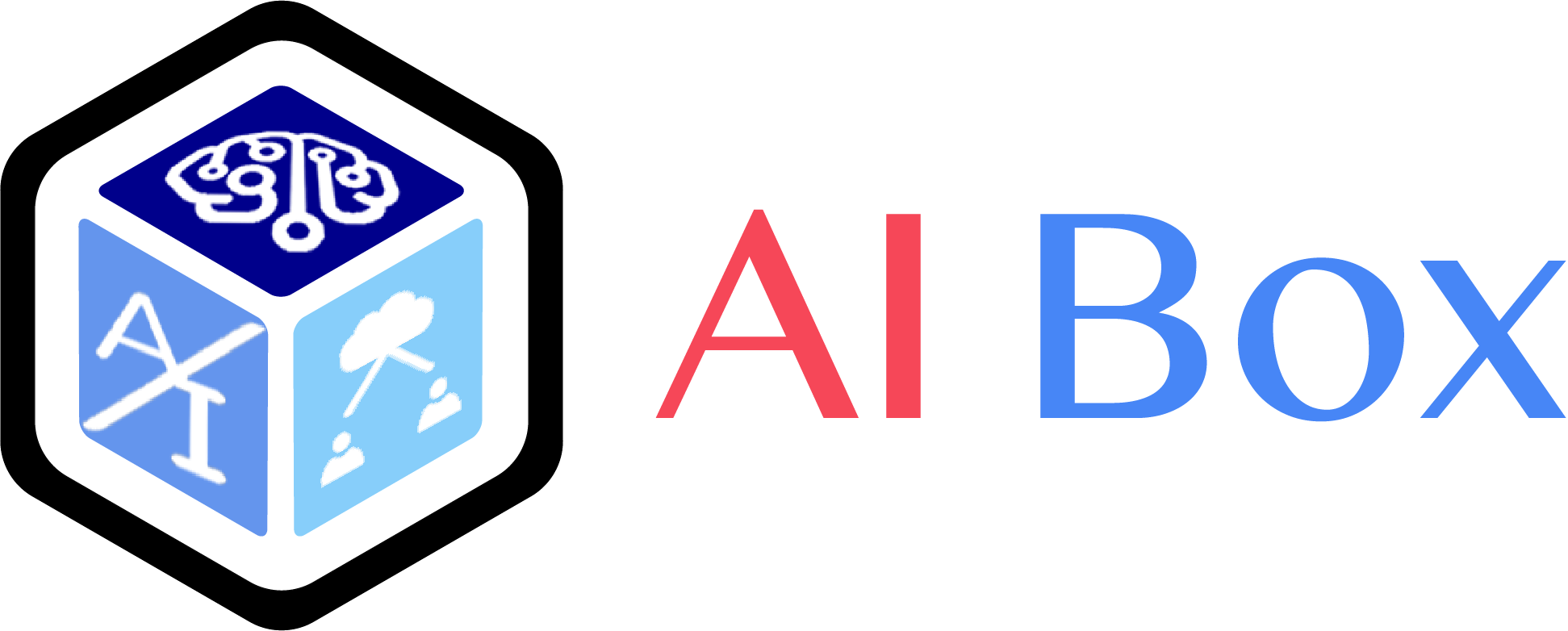}%
    \hfill
    \includegraphics[height=0.7cm]{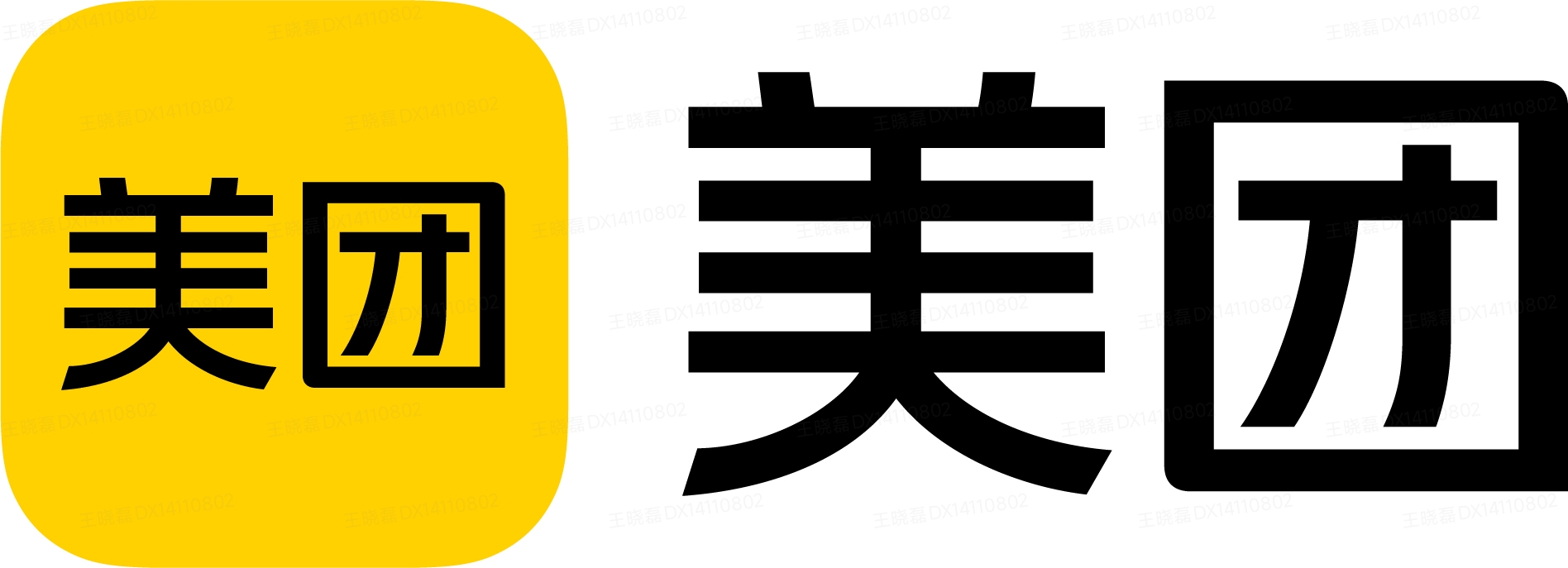}%
    }%
    \vspace{0.3em}
    \@toptitlebar
  }
  {}{}
\makeatother

\newcommand{\method}{RecPilot\xspace}

\title{Deep Research for Recommender Systems}

\author{%
  Kesha Ou$^{1}$\thanks{Equal contribution.}~, 
  Chenghao Wu$^{1*}$, 
  Xiaolei Wang$^{1*}$, 
  Bowen Zheng$^{1*}$, \\
  \textbf{Wayne Xin Zhao$^{1}$\thanks{Corresponding author.}~,} 
  \textbf{Weitao Li$^{2}$,}
  \textbf{Long Zhang$^{2}$,}
  \textbf{Sheng Chen$^{2}$,}
    \textbf{Ji-Rong Wen$^{1}$}
  \\
  $^1$Gaoling School of Artificial Intelligence, Renmin University of China, $^2$Meituan.\\
  \texttt{keishaou@gmail.com},
\texttt{batmanfly@gmail.com},
}

\begin{document}
\maketitle

\definecolor{babyblue}{HTML}{F8F9FE}
\newtcolorbox{bluebox}{
  colback=babyblue,    
  colframe=babyblue,  
  width=1.0\textwidth,  
  center,               
  arc=8pt,                 
  boxrule=0pt,           
  boxsep=0pt,       
  left=2pt,               
  right=2pt,              
  top=10pt,                
  bottom=10pt              
}


\begin{bluebox}
\begin{abstract}

The technical foundations of recommender systems have progressed from collaborative filtering to complex neural models and, more recently, large language models. 
Despite these technological advances, deployed systems often underserve their users by simply presenting a list of items, leaving the burden of exploration, comparison, and synthesis entirely on the user. This paper argues that this traditional ``tool-based'' paradigm fundamentally limits user experience, as the system acts as a passive filter rather than an active assistant.  
To address this limitation, we propose a novel deep research paradigm for recommendation, which replaces conventional item lists with comprehensive, user-centric reports. We instantiate this paradigm through \textbf{\method}, a multi-agent framework comprising two core components: a user trajectory simulation agent that autonomously explores the item space, and a self-evolving report generation agent that synthesizes the findings into a coherent, interpretable report tailored to support user decisions. This approach reframes recommendation as a proactive, agent-driven service. Extensive experiments on public datasets demonstrate that \method not only achieves strong performance in modeling user behaviors but also generates highly persuasive reports that substantially reduce user effort in item evaluation, validating the potential of this new interaction paradigm.

\end{abstract}
\end{bluebox}

\section{Introduction}
 

Recommender systems are designed to alleviate the problem of information overload by helping users discover relevant content from vast collections of items. To accurately capture and model user preferences, these systems have evolved through several key stages of technological advancement. 
Early approaches were primarily based on collaborative filtering (CF) techniques~\cite{koren2021advances}. With the rapid progress of deep learning, a variety of neural network architectures have since been widely adopted to enhance user representation learning and achieve more robust user–item matching.  These include multilayer perceptrons (MLP)~\cite{zhou2022filter}, recurrent neural networks (RNN)~\cite{DBLP:journals/corr/HidasiKBT15}, convolutional neural networks (CNN)~\cite{sharma2024comprehensive}, and the Transformer model~\cite{kang2018self}.
More recently, the emergence of large language models (LLMs)~\cite{llm_survey} has opened up new directions for development, enabling generative recommendation paradigms~\cite{tiger}, LLM-based recommendation frameworks~\cite{wu2024survey}, and the development of intelligent recommender agents~\cite{zhu2025recommender}.

Driven by these advances, user experiences in recommender systems have improved considerably in recent years. Looking ahead to future technical trends, however, it is worth asking a more fundamental question: Have existing recommender systems matured sufficiently, and have user experiences been fully maximized through such systems? To explore this question, we shift our perspective—setting aside our roles as recommendation algorithm researchers—and instead adopt the viewpoint of individual users interacting with these systems.  
From this vantage point, our firsthand experiences, likely shared by many users, suggest a more cautious answer. Consider e-commerce platforms as an illustrative example. In such environments, the dominant user intent is typically to locate and purchase a product that aligns with specific needs. To achieve this goal within existing recommender systems, users must browse a list of exposed items, click through potentially relevant ones to examine their details, and synthesize the accumulated information before making a final purchase decision. Despite technological advances, users remain underserved in this process: selecting items is still a labor-intensive endeavor, a burden is especially onerous for high-priced goods.

Upon closer examination of this negative finding, we identify a core limitation rooted in the position of recommender systems: the prevailing interaction paradigm of recommender systems remains focused on exposing relevant items. While this approach may have been reasonable in the early stages of recommender system design, it risks becoming a constraint on further improvements as technology evolves. Specifically, this paradigm may have ossified into a prevailing mindset that narrowly shapes how we think about and design recommender systems—a mindset that presupposes the necessity of active user participation in the decision process. It implicitly assumes, for instance, that users will compare presented items and scrutinize their details before making a final choice. 
To be more specific, existing recommender systems function more as \emph{tools} than as \emph{assistants}—that is, they facilitate access to information rather than orchestrating the complete recommendation process in a way that directly satisfies users' underlying intents, or at least significantly reduces users' effort in doing so.

To rethink the next-generation paradigm of recommender systems, we draw inspiration from the recent success of \emph{deep research}~\cite{openai_dr} in information retrieval, given that search and recommendation represent two highly related approaches to information seeking.
 In traditional search systems, users are presented with numerous web pages and must read through them individually to gather relevant information. By contrast, deep research systems employ an agent to autonomously interact with search engines, collect information from retrieved web pages on behalf of users, and synthesize the results into a comprehensive report for presentation. From a technical perspective, this information synthesis approach can be similarly applied to recommender systems. 
Specifically, the system—rather than the user—can undertake the labor-intensive process of item exploration, fulfilling a role analogous to that of the search engine in existing deep research systems.
Moreover, instead of simply generating a list of exposed items, the system could produce a more comprehensive report that directly presents the most relevant items alongside all necessary information to satisfy the user's needs. This approach would transform recommender systems from passive, user-driven filtering into proactive, agent-driven services, substantially reducing cognitive load and interaction costs while delivering explainable and trustworthy decisions through synthesized reports.


To this end, we propose a \emph{deep research} paradigm for recommender systems---a novel interaction approach that replaces conventional item lists with a comprehensive report. To realize this paradigm, we introduce \textbf{\method}, a multi-agent framework that explores the item pool on behalf of users and generates a research-oriented report to support decision-making. 
\method consists of two core components: a \textit{user trajectory simulation agent} and a \textit{self-evolving report generation agent}. The user trajectory simulation agent takes contextual information and historical user behaviors as input and actively simulates exploration trajectories within the item pool. We formulate this simulation as a generative modeling problem and employ reinforcement learning with model-free process rewards to enhance generalization. Leveraging the simulated trajectory, the report generation agent produces an interpretable report tailored to support user decisions. In doing so, the agent decomposes user interests into multiple aspects that users may consider during decision-making, generates a ranked list for each aspect, and integrates them into a comprehensive ranked list to facilitate rapid and informed choices.
To ensure personalization, we characterize user preferences using structured rubrics and experience-based memories, and introduce a self-evolving optimization mechanism that refines report generation along these two dimensions over time.

We evaluate the effectiveness of our approach on public recommendation datasets.
Experimental results demonstrate that our approach achieves superior performance in modeling observed user behaviors, with up to a 52\% improvement in Recall@5, verifying its reliability in exploring item candidates on behalf of users.
In addition, the generated reports exhibit excellent quality across multiple dimensions.
In particular, they often provide novel item recommendations that go beyond superficial preference matching (in 77\% of cases compared with the best baseline), which underscores its unique value in uncovering potential user preferences.

We summarize our main contributions as follows:

$\bullet$ To the best of our knowledge, this work represents the first attempt to fundamentally transform the interaction interface of recommender systems—moving from traditional item-centric lists to a user-centric, decision-support report.


$\bullet$ To instantiate this novel paradigm, we propose a multi-agent framework that autonomously explores the item pool on behalf of users and synthesizes the findings into a coherent, interpretable report designed to facilitate informed decision-making.


$\bullet$ {Extensive experiments on public recommendation datasets validate the effectiveness of our approach. In addition to achieving strong performance in modeling observed user behaviors, we show that the generated reports not only exhibit excellent accuracy and coverage, but also substantially reduce the user effort required for item comparison and evaluation. In particular, our reports can provide novel item recommendations that go beyond superficial preference matching, demonstrating the unique value of the deep research paradigm in recommender systems.
}
\section{Approach}

In this section, we formally define the deep research task and present our proposed multi-agent approach to address it.


\subsection{Overview of the Approach}
\label{sec:overview}

\paratitle{Task Formulation.}
\label{sec:task-form}
Deep research for recommender systems aims to reduce users' decision-making costs on e-commerce platforms. Specifically, we seek to minimize the cognitive and operational burden incurred from the moment a user enters the platform to the point of purchasing a satisfactory item. 
To achieve this goal, we focus on two sub-tasks: \emph{user exploration simulation} and \emph{exploration report generation}. 
Formally, we represent a user's historical behaviors as $X=[(a_1, v_1), \dots, (a_t, v_t)]$, where each tuple $(a_i, v_i)$ denotes an item-behavior pair. The exploration process on the platform is modeled as a consecutive behavior session $Y = [(a_{t+1}, v_{t+1}), \dots, (a_{t+T}, v_{t+T})]$, with $T$ being the number of behaviors occurring during exploration. 
User exploration simulation aims to emulate an efficient traversal of the item pool, formalized as a conditional generation task $g_{\text{exp}}(Y \mid X)$\footnote{Here we slightly abuse the probabilistic notation of conditional ``$\mid$'' to denote the involved context.}, where $g_{\text{exp}}$ is the exploration function that generates items a user might find highly relevant. 
Based on the simulated exploration trajectory, we generate a research report $\mathcal{R}$, formalized as $\mathcal{R} \sim f_{\text{rep}}(\cdot \mid X, Y)$, where $f_{\text{rep}}(\cdot)$ denotes a report generation function that can be instantiated through various architectural designs. The resulting report augments the recommended item set $Y$ with comprehensive explanations and justifications, all grounded in the user's historical behavior context $X$, thereby facilitating more informed and confident decision-making.



\begin{figure}[t]
    \centering
    \includegraphics[width=\linewidth]{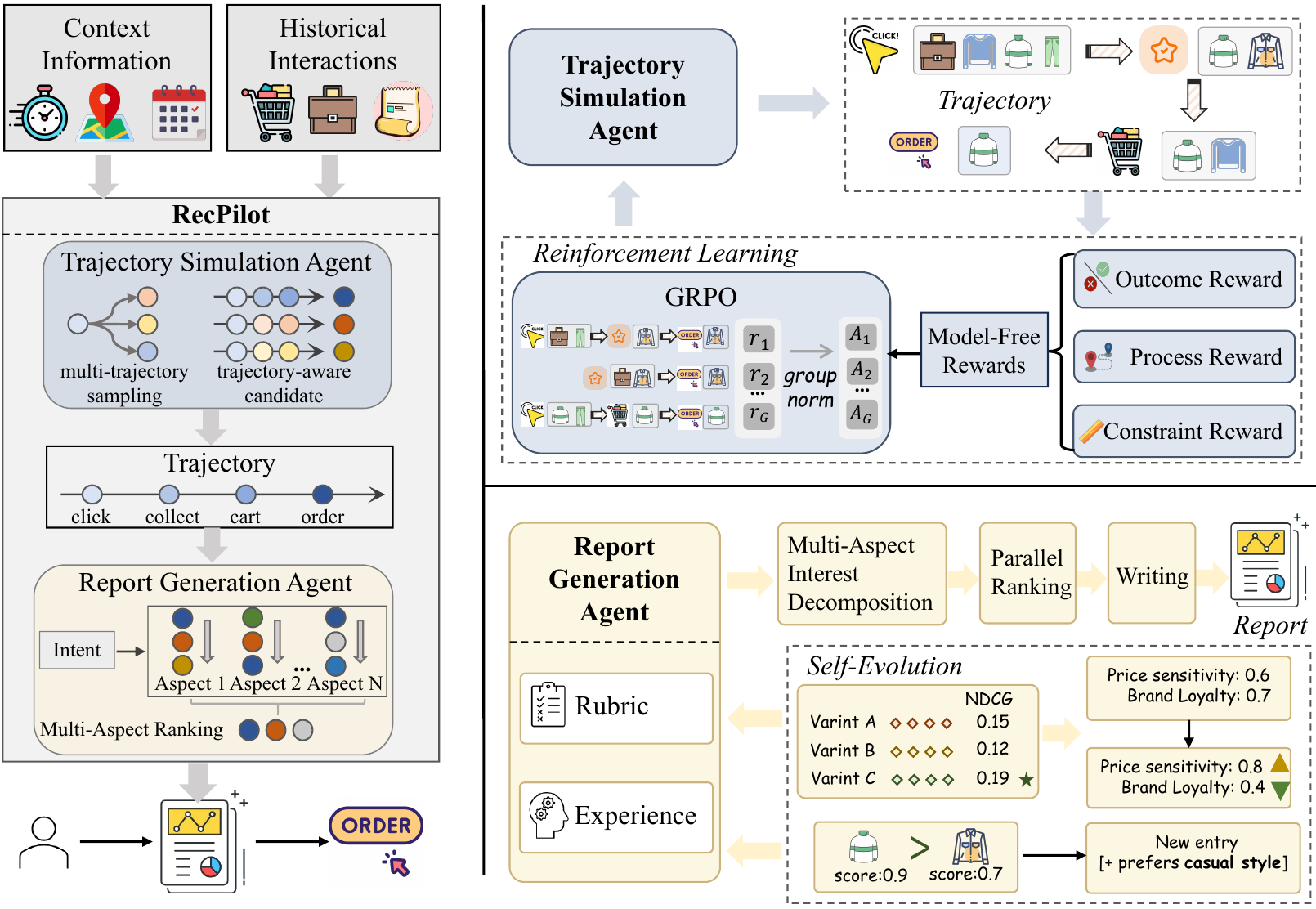}
    \caption{The overview of our approach \method. The left part demonstrates the overall pipeline. The trajectory simulation agent receives contextual information and historical interactions as input, and generate a simulated trajectory for the report generation agent to generate the final report. 
    The upper-right part introduces the trajectory agent, which interleaves item and behavior generation and is optimized via model-free reward reinforcement learning. The lower-right part illustrates the report generation agent, which first decomposes user interests into multiple aspects for ranking before writing reports. It characterizes user preferences through rubrics and experience, and achieves self-evolution along these two dimensions for optimization.}
    \label{fig:model}
\end{figure}

\paratitle{Overall Architecture.}
In this paper, we propose \textbf{\method}, a multi-agent framework that explores the item pool on behalf of users and generates an exploration report to assist them in making final purchase decisions. 
\method comprises two core agents: a \textit{user trajectory simulation} agent (Section~\ref{sec:simulation-agent}) and a \textit{self-evolving report generation} agent (Section~\ref{sec:report-agent}), which instantiate the exploration function $g_{\text{exp}}(\cdot)$ and the report generation function $f_{\text{rep}}(\cdot)$, respectively. 
As illustrated in Figure~\ref{fig:model}, the user trajectory simulation agent takes contextual information and historical user behaviors as input, and proactively simulates exploration behaviors within the item pool. By offloading the initial low-value and repetitive interaction processes from users, this agent effectively alleviates the burden of tedious exploration. To model this simulation task, we adopt a generative approach and employ reinforcement learning with model-free process rewards, enabling more robust recommendation performance under complex contextual conditions. 
The simulated trajectory is then passed to the report generation agent, which transforms it into an interpretable report designed to support decision-making. In this process, the agent decomposes user interests into multiple aspects that users are likely to consider during evaluation, generates a ranked list for each aspect, and subsequently integrates them into a comprehensive ranked list to enable rapid and informed decisions. To ensure personalization, we characterize user preferences using structured rubrics and experience-based memories, and introduce a self-evolving optimization mechanism that refines report generation along these two dimensions over time.

\subsection{User Trajectory Simulation Agent}
\label{sec:simulation-agent}

In large-scale e-commerce scenarios, users often face high time costs and exploration fatigue, which hinders their ability to efficiently discover items of potential interest. To address this challenge, we propose a \emph{user trajectory simulation agent}. By proactively simulating user exploration behaviors within the candidate space, this agent acts as a surrogate that performs deep, targeted, and fine-grained interactions—inferring the items a user is likely to prefer—thereby relieving users from the burden of manual, repetitive exploration. 
Specifically, the agent first learns and extracts users' dynamic interests and preferences from contextual information and historical behavior sequences. It then adopts an autoregressive generative paradigm to synthesize high-quality, diverse, and length-adaptive \emph{exploration-to-decision} trajectories. Through this behavioral path simulation, the agent effectively eliminates the need for users to manually sift through candidate items. More importantly, these simulated trajectories serve as explicit reasoning pathways that facilitate deeper cognitive processing within the agent, enabling more accurate retrieval of relevant candidates. As a result, the agent delivers a high-quality candidate set to support downstream decision-making modules.

\subsubsection{Generative User Trajectory Learning}

This section presents a generative framework for learning user trajectories. We first reconstruct the interaction data into a session-aware sequential format. Building on this formulation, we then employ supervised learning (SL) to guide the model in acquiring the exploration-to-decision paradigm.


\paratitle{Session-Aware Trajectory Tokenization.}
In e-commerce scenarios, user decision-making typically unfolds along a progressive trajectory from exploration to decision. The final purchase decision exhibits a strong causal dependency on preceding exploratory behaviors, and different behavior types naturally exhibit hierarchical levels of engagement (e.g., $\texttt{purchase} > \texttt{favorite} > \texttt{click}$).
Unlike conventional approaches that flatten the actions and items in a trajectory into independent tokens~\cite{mbgen}, we adopt an action-guided aggregation scheme to better structure user trajectories by organizing action–item interactions into prefix-structured segments.
To achieve this, we first map both heterogeneous action types and discrete item IDs into a unified vocabulary space. 
Then, for consecutive behaviors with the same action type, we take the corresponding action token as the prefix, followed by all the item tokens arranged in chronological order.
Recall that the trajectory is refered to a consecutive behavior session (Section~\ref{sec:overview}), thus, to further enhance the awareness of session boundaries, we prepend and append special tokens \texttt{<bos>} and \texttt{<eos>} to each trajectory.
Formally, a tokenized trajectory is represented as:
\begin{equation}
\tilde{S} = h(S) = \langle \texttt{<bos>}, a_{[1,2]}, v_1, v_2, a_{[3,4]},v_3,v_4, \cdots, a_{T}, v_T, \texttt{<eos>} \rangle,
\end{equation}
where $S=[(a_1, v_1), \dots, (a_T, v_T)]$ is the original user session comprising sequential action-item pairs (as defined in Section~\ref{sec:task-form}), $h(\cdot)$ denotes the proposed tokenization function, $a_{[i,j]} = \{ \texttt{<click>}, \texttt{<purchase>}, \texttt{<favorite>}, \texttt{<cart>} \}$ represents the action token covering the consecutive $i$-th to $j$-th interactions of the same type, and $a_{T}$ is fixed as the $\texttt{<purchase>}$ action, indicating the completion of the user's decision-making process and the termination of the current session. 
By collapsing contiguous interactions of the same type, this format substantially compresses high-frequency $\texttt{<click>}$ tokens. Such compression mitigates the interference caused by long-tail action distributions, encouraging the model to focus on inherent state transitions rather than merely replicating frequent action tokens.


\paratitle{Multi-Behavior Trajectory Modeling.}
\label{sec:trajectory model}
We formulate the modeling of user trajectories as a sequence-to-sequence generation task. 
Given a tokenized user historical sequence, $\tilde{X} = h(X)$, the model is trained to autoregressively generate the subsequent ``exploration-to-decision'' trajectory (\ie a sequence of exploratory behaviors such as clicks and favorites culminating in a final purchase), which is denoted as $\tilde{Y} = h(Y)$.
To address the inherent class imbalance within historical interaction data (where frequent click actions significantly outnumber sparse purchase actions), we employ an encoder-decoder architecture (\eg T5~\cite{t5}). 
By decoupling the encoding of historical context from the generation of future actions, we prevent the decoder from merely over-imitating historical patterns or biasing predictions toward high-frequency behaviors. 
During this stage, we utilize supervised learning to optimize the model, establishing a foundational capability to simulate user exploration trajectories.
The optimization objective is as follows:
\begin{align}
    \mathcal{L} = - \sum_{\langle \tilde{X},\tilde{Y} \rangle \in \mathcal{B}} \sum_{t=1}^{|\tilde{Y}|}\operatorname{log}P(\tilde{Y}_t|\tilde{X},\tilde{Y}_{<t}),
\end{align}
where $\mathcal{B}$ denotes the training set.
$\tilde{Y}_t$ is the $t$-th token of target session $\tilde{Y}$ and $\tilde{Y}_{<t}$ denotes the tokens before $\tilde{Y}_t$.

\subsubsection{Reinforcement Learning with Model-Free Process Rewards}

Supervised learning (SL) enables the model to initially capture the behavioral evolution patterns of users from exploration to decision.
However, as a teacher-forcing paradigm, it strictly replicates patterns observed in the training data, which often results in limited generalization and inadequate exploration capabilities.
This often results in limited generalization and a reduced capacity for stochastic exploration. 
To address these limitations, we introduce a reinforcement learning (RL) phase to further improve the model's generalization and exploratory performance.
Specifically, we generate multiple exploration trajectories via stochastic sampling and incorporate model-free process rewards to discriminate and evaluate their quality.
Finally, we employ the GRPO algorithm~\cite{grpo} to optimize the stability and quality of the generated trajectories during long-term reasoning.

\paratitle{Model-Free Process Rewards.}
The reward function serves as the primary training objective in RL. 
Recent studies have highlighted the advantages of rule-based rewards~\cite{deepseek-r1, kimi-k1.5}, which are easy to develop and hard to hack.
In light of these benefits, our approach adopts a similar model-free, rule-based reward approach. 
To jointly optimize both the exploratory simulation process and the final recommendation quality, we propose a composite reward function comprising three distinct components: outcome reward, process reward, and constraint reward.

$\bullet$ \textit{Outcome Reward:}
To enhance recommendation performance, we introduce an outcome reward to evaluate whether the generated trajectory leads to the correct decisions. 
Specifically, the model receives a positive reward when the final predicted item $\hat{v}_T$ matches the actually purchased item $v_T$, formulated as:
\begin{align}
\label{equ:or}
   R_O = \mathbb{I}(\hat{v}_T = v_T), 
\end{align}
where $\mathbb{I}(\cdot)$ denotes the indicator function.
This reward directly aligns with the ultimate objective of the recommender system, offering a clear optimization target for reinforcement learning. However, the reward is inherently sparse, making it difficult to provide effective guidance for intermediate exploration behaviors.



$\bullet$ \emph{Process Reward:}
Relying solely on the outcome reward insufficient to capture the quality differences between different exploration trajectories during the generation process.
Therefore, we further introduce process rewards to measure the semantic consistency between generated exploration trajectories and the ground truth user trajectory. 
This design is motivated by a key observation in e-commerce scenarios: effective user exploration simulation does not require token-by-token replication of historical interactions. Instead, it should preserve consistency in the semantics and collaborative relationships of the browsed items. Strict ID-level matching risks suppressing the natural diversity of user behaviors and hinders the model's ability to generalize to plausible yet unseen trajectories. 
Based on this insight, we adopt a process reward grounded in collaborative consistency, moving beyond hard matching. 
Specifically, we leverage the ID embedding learned during the SL phase for defining a collaborative semantic space. 
For each valid item token in the generated trajectory, we pair it with every valid item token in the ground truth trajectory, compute the cosine similarity for each pair, and select the maximum similarity score as the individual reward for that generated item, followed by a mean-pooling step across all generated item tokens. The overall process reward is thus formulated via Max-Sim pooling~\cite{colbert}:

\begin{align}
\label{equ:pr}
R_P = \frac{1}{| \mathcal{V}_{\hat{Y}} |} \sum_{\hat{v} \in \mathcal{V}_{\hat{Y}}}  \max_{v \in \mathcal{V}_Y} 
\frac{\bm{e}_{\hat{v}} \cdot \bm{e}_{v}}{\|\bm{e}_{\hat{v}}\| \, \|\bm{e}_{v}\|},
\end{align}
where $\mathcal{V}_{\hat{Y}}$ denotes the items in the generated trajectory $\hat{Y}$, and $\mathcal{V}_{Y}$ represents the items in the ground truth trajectory $Y$. 
This design is justified from two perspectives. 
First, 
the strict ordering among items within the same interaction stage (\eg consecutive clicks) is often weak and can be easily influenced by external factors, such as page layout or recommendation positions. 
Second, by encouraging semantic alignment rather than exact item-ID matching, the model learns to generate reasonable and diverse exploration trajectories, thereby improving its generalization capability to unseen user behavior patterns.

$\bullet$ \emph{Constraint Reward:}
To further optimize trajectory generation, we introduce lightweight constraint rewards to regularize both trajectory length and format. 
The length constraint guides the model to generate trajectories whose length aligns with the user's actual exploration depth. 
It penalizes deviation between the generated trajectory length $L_{\hat{Y}}$ and the ground-truth length $L_{Y}$ via an exponential decay function:
\begin{align}
R_L = e^{-\mu |L_{\hat{Y}} - L_{Y}|},
\end{align}
where $\mu$ is a scaling factor set to 0.2 here. 
As previously discussed in Section~\ref{sec:trajectory model}, a valid trajectory must adhere to the predefined ``exploration-to-decision'' sequence. To strictly satisfy this structural requirement, we introduce a format penalty $R_F$ to constrain the generation process via the following two hard rules. A trajectory is deemed incorrectly formatted if: (1) it begins directly with a \texttt{<purchase>} action (i.e., zero exploration steps), or (2) it contains consecutive identical action tokens without intervening items, rendering the step meaningless for actual item exploration. 
The corresponding penalty is defined as:
\begin{align}
\label{equ:cr-format}
R_F = \begin{cases}
0 & \text{if the format is correct,} \\
-1 & \text{otherwise.}
\end{cases}
\end{align}
The final composite constraint reward is the sum of the two components:
\begin{align}
    \label{equ:cr}
  R_C =  R_L +  R_F.
\end{align}

\paratitle{Policy Update via GRPO.}
For each user, we generate a group of $G$ trajectories, denoted as ${o_1, \dots, o_G}$, by sampling from the current policy $\pi_\theta$. The total reward associated with a trajectory $o_i$ integrates the aforementioned three rewards as follows:
\begin{equation}
r_i = R_O + R_P + R_C,
\end{equation}
where $R_O$ (Eq.~\eqref{equ:or}) is the outcome reward evaluating the final prediction, $R_P$ (Eq.~\eqref{equ:pr}) is the process reward assessing the intermediate reasoning, and $R_C$ (Eq.~\eqref{equ:cr}) is the constraint reward enforcing formatting rules.
To reduce variance without introducing a separate value network, GRPO estimates the advantage $A_i$ for each trajectory by normalizing rewards within the group:
\begin{equation}
A_i = \frac{r_i - \text{mean}({r_1, \dots, r_G})}{\text{std}({r_1, \dots, r_G})}.
\end{equation}
The policy $\pi_\theta$ is then updated by maximizing the clipped objective:
\begin{equation}
    L_{\text{GRPO}} 
= \frac{1}{G} \sum_{i=1}^G \Big( \min ( r_i A_i, \, \text{clip}(r_i, 1-\epsilon, 1+\epsilon) A_i ) -\beta \, \mathbb{D}_{\text{KL}}(\pi_\theta \,||\, \pi_{\text{ref}}) \Big),
\end{equation}
where $r_i = \pi_\theta(o_i) / \pi_{\text{ref}}(o_i)$, and $\pi_{\theta}$ and $\pi_{\text{ref}}$ denote the current policy model and the reference model, respectively. The term $\mathbb{D}_{\mathrm{KL}}$ is a KL regularization that prevents the policy from deviating excessively from the initial model distribution.

\subsubsection{Exploration Trajectory Generation}
During inference and generation, the model acts as a user agent, performing simulated exploration within a large-scale item space. 
This simulation yields a diverse set of exploration-to-decision trajectories, ranging from broad, divergent browsing to progressively focused intent convergence. Specifically, the inference process comprises two core components.

\paratitle{Diverse Trajectory Sampling for Interest Exploration.}
Due to the randomness and inherent uncertainty of user interaction behaviors, greedy decoding strategies often fail to capture the full distribution of potential user intentions. While beam search can generate multiple candidate paths, the resulting trajectories typically exhibit only minor deviations, particularly in their final segments. This limitation hinders the model's ability to represent the diverse branches of user interest and leads to insufficient exploration of the item space.
To better simulate the evolution of user intentions, we employ a multi-trajectory sampling approach during decoding. Specifically, we adopt a top-$p$ sampling strategy, where at each step we sample from tokens whose cumulative probability mass exceeds a threshold $p$. This allows us to generate $N$ distinct exploration trajectories in parallel, capturing a broader range of plausible user behavior paths.


\paratitle{Trajectory-Aware Candidate Generation.}
\label{sec:candidate-generation}
Based on the $N$ exploration trajectories generated above, we infer the user's intention convergence point and predict the target item most likely to trigger a decision action (\ie purchase).
For each generated exploration trajectory, we utilize the hidden state from its final step to predict the next item, retrieving the top-$K$ relevant candidates. 
This procedure yields an initial candidate pool of size $N \times K$.
Subsequently, we concatenate the exploration trajectory with final predicted item to form a complete simulated trajectory. The log-likelihood of this trajectory serves as the final score for the candidate item.
Finally, after removing duplicate items from the candidate pool, we select the top-$K$ items with the highest scores as the final retrieval results, and retain their associated exploration trajectories—\ie  the ``exploration-to-decision'' trajectories. These retained trajectories serve as high-confidence simulations of user behavior and are subsequently passed to the downstream report generation agent for further intention analysis and utilization.



\subsection{Self-Evolving Report Generation Agent}
\label{sec:report-agent}
In this section, we present the development of the report generation agent, which aims to transform the simulated trajectory into a human-readable report for decision support. Next, we detail how to generate the report and improve its quality through self-evolution.

\subsubsection{Agentic Report Generation}
\label{sec:report}

As described in Section~\ref{sec:candidate-generation}, the user trajectory simulation agent generates multiple candidate items that a user may find relevant. However, presenting all candidates directly in the report would still impose a high cognitive load on the user. Therefore, it is necessary to first select the most promising items from this candidate set. 
This selection process is analogous to the fine-ranking stage in conventional recommendation systems, where each candidate item is assigned a score to determine the final ranked list. Drawing inspiration from this paradigm, we perform a final ranking over the candidate items produced by the trajectory simulation agent. Crucially, this ranking should be informed by a comprehensive consideration of the various dimensions of user preference. 
To achieve this, we first decompose user interest into multiple aspects (\ie subsets of attributes). Then, we perform ranking independently for each aspect and subsequently integrate the aspect-wise rankings to produce a consolidated final list. Once the ranked list is obtained, we transform it into a human-readable report. This report not only presents the top-ranked candidate items both within and across aspects but also supplies detailed interpretive information, structured in a narrative format, to support informed user decision-making.

\paratitle{Rubric-Experience Dual-Channel Preference.}
To perform effective ranking and support interpretable decision guides in the report, we propose to characterize user preferences from two perspectives, \ie structured rubrics and textual experience.
Structured rubrics represent the skeleton of user preferences.
To support numerical comparison among items, we utilize item attributes as common rubrics across users.
For each user $u$, the preference is represented as a set of priority scores over each attribute $a \in \mathcal{A}$, which can be denoted as $\mathcal{W}_u = \{ w_u^{(a)} \mid a \in \mathcal{A} \}$.
Complementing structured rubrics, textual experience captures the semantic nuances of user preferences.
Since attribute-based rubrics struggle to encode context-dependent preferences, textual experience serves as a flexible and semantically rich supplement to capture implicit, scenario-specific user inclinations.
For each user $u$, we maintain an experience memory $\mathcal{E}_u$, where each experience entry is structured as a key-value pair, with contextual conditions serving as its key and the relevant content as its value.

\paratitle{Multi-Aspect Interest Decomposition For Parallel Ranking.}
To support the aspect-aware decision guide part in the report, we first decompose user interest into multiple aspects based on the rubrics and experience introduced in the above part and then perform parallel ranking against each specific aspect.
Here, we associate an aspect of user preference with a subset of attributes, which can be denoted as $\mathcal{D} = \{ a \mid a \in \mathcal{A} \}$.
The importance weight of each attribute $a$ for a user $u$ correspond to its rubric score $w_u^{(a)}$ defined in the above part.
To decompose user interest into multiple aspects, we propose an LLM-based approach.
Specifically, for a target user $u$, we first prompt an LLM to compress the simulated trajectory into an intent summary.
Then, we utilize the intent summary as a query $q$ to retrieve a certain number of experience from the experience memory $\mathcal{E}_u$.
This retrieval process is performed by computing the semantic similarity between the intent summary and the contextual conditions of each entry in $\mathcal{E}_u$, followed by selecting the most similar entries.
{Taking the intent summary and retrieved experiences as context, we leverage the LLM to identify multiple attribute subsets, which formally constitute the decomposed aspects of user preference.}
For each aspect, we take the corresponding rubric scores of associated attributes and increase each entry by $\delta$ to emphasize their relevance.
Subsequently, we perform ranking for each aspect in parallel.
Here, we take one aspect $\mathcal{D}$ as example to detail the ranking process.
For each candidate item $i$, we first prompt an LLM to evaluate the matching degree with each attribute $a$ associated with the aspect and generate a matching score $s_i^{(a)}$.
Next, we apply rubric scores for each attribute to calculate a weighted sum of all the attribute matching scores. 
This yields the final item score for aspect $\mathcal{D}$, expressed as follows:
\begin{equation}
    s_i^{(\mathcal{D})} = \frac{1}{|\mathcal{D}|}\sum_{a \in \mathcal{D}}{w_u^{(a)} \cdot s_i^{(a)}}.
\end{equation}

In this way, we complete the ranking for one aspect.
To derive the ranked list across various aspects, we directly perform a summation operation over all the scores across all the aspects as the final score, and sort items accordingly.

\paratitle{Structured Report Generation.}
After obtaining the ranked list within each aspect and across aspects, we extend it into a human-readable report with detailed interpretive information for decision support.
We organize the report $\mathcal{R}$ in a structured manner, consisting of four components: simulated trajectory, user intent summary, the overall item suggestions, and the item suggestions for each aspect.
It can be represented as follows:
\begin{equation}
\label{eq:report}
    \mathcal{R} = \left\{ h(\hat{Y}), h(q), h(L_0), h(L_1, \dots, L_{n}) \right\},
\end{equation}
where $h(\cdot)$ denotes a function that transforms input content into a report component using an LLM, $\hat{Y}$ represents the simulated trajectory, $q$ denotes the user intent summarized from the trajectory, $L_0$ is the overall ranked list across all aspects, and $L_1, \dots, L_{n}$ are the ranked lists corresponding to each aspect.
For the last component, we leverage the attributes in each aspect to generate corresponding rationales, which serve as the guidance to support decision-making.

\subsubsection{Self-Evolution for Personalization}
\label{sec:evolution}



In the preceding section, we discussed the approach to generating human-readable reports. 
However, a critical challenge in real-world recommendation scenarios is that user interests are not static; 
they dynamically evolve and drift over time across multiple interactions. 
Consequently, preferences captured during the initial exploration phase may gradually become outdated. 
To provide consistently accurate and personalized decision support, the recommender agent must be capable of continuously learning and adapting to these dynamic shifts, which necessitates an evolution mechanism.
Furthermore, to address shifts in user interest, traditional recommender systems typically require the manual collection of new data and periodic retraining or fine-tuning of the underlying models. 
To accelarate the evolution, we introduce a training-free self-evolution mechanism. 
Rather than relying on manual model updates, our system enables the agent to autonomously refine the structured rubrics and experience memories of a user (as defined in Section~\ref{sec:report}) in a real-time, closed-loop manner. 
By continuously gathering and assimilating ongoing user feedback during service, this self-evolution mechanism ensures the timeliness and accuracy of personalized decision support without incurring any additional training overhead.

\paratitle{Rubric Optimization.}
To keep rubrics aligned with evolving user interests and mitigate the impact of noise, we propose a best-of-$n$ optimization method.
Recall that in Section~\ref{sec:report}, we decompose user interests into multiple aspects and generate corresponding ranked lists.
To perform optimization, we first evaluate each ranked list against high-level user behaviors\footnote{High- and low-level behaviors are distinguished based on their frequency of occurrence, where high-level behaviors typically serve as stronger signals of user intent.
} (\ie purchases) using the NDCG metric~\cite{jarvelin2002cumulated}. We then identify the list that yields the highest NDCG score and employ its associated rubric scores (which are increased by $\delta$) to update the original values. 
By selecting only the top-performing variant, we effectively filter out suboptimal or noisy rubric adjustments, ensuring that the updated scores strictly align with high-level behavioral preferences.

\paratitle{Experience Consolidation.}
To keep the experience memory aligned with evolving user interests, we propose a contrastive strategy that extracts user preference cues and incorporates them as new experience entries. Specifically, we take the ranked list that achieves the highest NDCG score during rubric optimization as the positive sample, while using the ranked list generated according to the rubric scores before optimization as the negative sample. When items associated with high-level behaviors are ranked higher in the positive example, we construct a contrastive pair using the two ranked lists and prompt an LLM to generate corrective experience entries that explicitly articulate the decision logic requiring adjustment under specific scenarios. This contrastive approach amplifies the discrepancies between the suboptimal original ranked list and the optimized high-performance one, enabling targeted and interpretable refinement of the experience memory.


\paratitle{Extended Experience Mining.}
The optimization described above relies on user sessions that contain high-level behaviors (\eg purchases). However, such sessions are scarce in real-world scenarios. To address this data scarcity, we further incorporate sessions with low-level behaviors (\eg clicks) and adopt an LLM-based mining method to enrich preference learning. 
Specifically, we prompt an LLM to perform reasoning over each low-level session, inferring potential negative preferences or unmet needs that can supplement the existing experience base. Notably, we do not use these sessions to update rubrics, as rubrics encode user preferences in precise numerical form, and low-level behaviors are inherently noisy.

\section{Experiment}

\subsection{Experimental Setup}
\label{sec:setup}

\paratitle{Dataset.}
We evaluate our model on the \textsc{Tmall} dataset\footnote{https://tianchi.aliyun.com/dataset/140281}, which consists of four kinds of user behaviors: \emph{click}, \emph{collect}, \emph{add to cart}, and \emph{purchase}.
We take ``purchase'' as the target behavior to be predicted, while treating other behaviors as part of the exploratory process preceding the decision.
We select two weeks of interaction records and segment user sessions on a daily basis. 
To mitigate data sparsity, we filter out items that appear fewer than five times in the training set.
The processed data comprises 288,777 users, 556,233 items, and 1,294,696 sessions.
For dataset splitting, we adopt the session-wise leave-one-out strategy following existing work~\cite{DBLP:journals/corr/abs-2511-03155}.
Since our objective is to predict purchases, we exclusively retain sessions containing purchase behaviors for validation. This results in a final split of 306,088 sessions for training, 1,941 for validation, and 1,272 for testing.

\paratitle{Baselines.}
In this paper, we consider two sub-tasks: user exploration simulation and exploration report generation, as stated in Section~\ref{sec:overview}.
For comparison, we select several representative methods tailored for each subtask.

(1) {Sequential recommendation methods}:

$\bullet$ \textbf{SASRec}~\cite{sasrec} employs a self-attention mechanism to capture long-term dependencies within user interaction sequences, configured with 2 layers and 2 attention heads. 

$\bullet$ \textbf{BERT4Rec}~\cite{bert4rec} utilizes a bidirectional Transformer architecture with the Cloze task to learn deep behavior representations, implemented with 2 layers, 2 heads, and a masking ratio of 0.2.

$\bullet$ \textbf{GRU4Rec}~\cite{gru4rec} uses a Gated Recurring Unit (GRU) to enhance long-term memory in recommendation tasks, utilizing 2 recurrent layers with a dropout rate of 0.1.

(2) Multi-behavior sequential recommendation methods:

$\bullet$ \textbf{PBAT}~\cite{su2023personalized} identifies personalized interaction signatures by integrating user-specific behavioral features, parameterized by 2 layers and 2 attention heads.

$\bullet$ \textbf{MBHT}~\cite{yang2022multi} reveals hidden multi-behavior associations between items by modeling their interactions within a hypergraph structure, configured with 2 layers, 2 heads, a masking ratio of 0.2, and multi-scale hyperedges of [5, 8, 40].

$\bullet$ \textbf{MB-STR}~\cite{yuan2022multi} addresses the inconsistency across heterogeneous behaviors by employing a sparse MoE-based selection mechanism, implemented with 2 layers, 2 heads, 3 shared experts, and 32 routing buckets.

(3) Reasoning enhanced sequential recommendation methods:

$\bullet$ \textbf{ReaRec}~\cite{tang2025think} incorporates a latent reasoning stage before the final prediction to analyze the underlying motivations behind user histories, configured with 2 reasoning steps.

(4) {LLM-based report generation methods}:


We categorize these baselines into two groups. 
The first group is \textbf{direct reasoning}, which includes leading LLMs such as \texttt{DeepSeek-V3.2}~\cite{liu2025deepseek}, \texttt{GLM-5}~\cite{zeng2025glm}, \texttt{Qwen3-Max}~\cite{yang2025qwen3}, and \texttt{GPT-5.2}~\cite{singh2025openai}. 
These models receive the same input as our method and are directly instructed to generate a decision report. 
The second group comprises \textbf{agent-based methods}, which utilize agentic workflows to handle the complex generation task. This group includes:

$\bullet$ \textbf{ReAct}~\cite{react} interleaves reasoning with task-specific actions, allowing the model to dynamically adjust its thoughts based on observations.

$\bullet$ \textbf{Plan-and-Solve}~\cite{wang-etal-2023-plan} decomposes a complex task into manageable subtasks and solves them sequentially following a static predefined plan.





\paratitle{Evaluation Metrics.}
For the user exploration simulation subtask, as the target behavior is the most critical factor directly associated with the final ranked list in the report, we focus on its prediction accuracy.
Specifically, we employ Recall@$k$ and NDCG@$k$ ($k \in \{5, 10\}$) for performance evaluation.
For the exploration report generation subtask, since it is subjective to assess the quality of a report, we design six important demensions to make it measurable.
Each metric ranges from 1 to 5 points, with detailed descriptions as follows:

$\bullet$ \textbf{Accuracy}
measures whether the items selected after reviewing the report are the same as labels.
    
$\bullet$ \textbf{Coverage}
    measures whether the report captures and presents all the core attributes that are influential to the decision-making process.
    
$\bullet$ \textbf{Informativeness}
    quantifies the proportion of verifiable, factual content relative to unsubstantiated or speculative elements within the report.
    
$\bullet$ \textbf{Clarity}
    assesses whether the report delivers unambiguous comparisons and clear conclusions, thereby effectively supporting the user in making an informed decision.
    
$\bullet$ \textbf{Consistency}
    measures whether the user intent and item ranked lists parts strictly follows the simulated trajectory.
    
$\bullet$ \textbf{Novelty}
    evaluates whether the report proposes extra interests beyond simply matching the intent and whether such proposals are supported by traceable evidence.

To perform evaluation based on these subjective dimensions, we employ both humans and LLMs.
For human evaluation, we invite ten real users to conduct a blind test on randomly selected examples. To mitigate subjective bias and ensure reproducibility, we strictly follow a human evaluation protocol rather than relying on holistic impressions. Evaluators are required to score the generated reports independently based on a fine-grained 5-point scale with explicit behavioral anchors for each of the six dimensions. 
For LLM-based evaluation, we adopt \texttt{Gemini-3-flash}~\cite{team2023gemini} as simulated users and prompt it with the same scoring instruction.
The detailed evaluation guidelines, task setup, and the complete scoring rubric are provided in Appendix~\ref{app:human-evaluation}.





\paratitle{Implementation Details.}
For our user trajectory simulation agent, we employ T5~\cite{t5} as the backbone model. 
It is configured with 2 encoder layers and 2 decoder layers. 
The hidden dimension is set to 64, while the feed-forward network has an intermediate dimension of 256 and uses the ReLU activation function. For causal self-attention, we use 2 attention heads, each with a head dimension of 64.
We optimize the model using the AdamW optimizer with a batch size of 32. The initial learning rate is set to $1 \times 10^{-3}$, and follows a cosine annealing schedule with a linear warm-up phase.
The final model checkpoint is selected based on the NDCG@10 metric on the validation set.
For our self-evolving report generation agent, we adopt \texttt{DeepSeek-V3.2}~\cite{liu2025deepseek} as the backbone. 
To ensure a fair comparison, all methods are evaluated using the same decoding settings: temperature is set to 0.2 and the maximum number of tokens to generate is 16384. 
We employ \texttt{Qwen3-Embedding-8B}~\cite{zhang2025qwen3} as the text embedding model for experience retrieval. 
To mitigate the influence of formatting differences on evaluation, we standardize the output format across all approaches, requiring each generated report to include four components as Eq.~\eqref{eq:report}.
Our code is available at \url{https://github.com/RUCAIBox/RecPilot}.


\begin{table*}[t]
\caption{Performance comparison on the \textsc{Tmall} dataset about the trajectory simulation task. The best results are marked in bold.}
\label{tab:tmall_results}
\resizebox{\textwidth}{!}{
\begin{tabular}{lcccccccccc}  
\toprule
\multirow{2}{*}{Metric}  
& \multicolumn{3}{c}{Sequential} 
& \multicolumn{3}{c}{Multi-behavior} 
& \multicolumn{1}{c}{Reasoning} & \multirow{2}{*}{Ours} \\  
\cmidrule(lr){2-4} \cmidrule(lr){5-7} \cmidrule(lr){8-8}  
 & GRU4Rec & BERT4Rec & SAS4Rec 
 & PBAT  & MBHT & MBSTR & ReaRec &  \\ 
 \midrule
Recall@5  & 0.0618 & 0.0527 & 0.0636 & 0.0551 & 0.1015 & 0.1025 &  0.0794 &\textbf{0.1557} \\
Recall@10 & 0.0793 & 0.0708 & 0.0808 & 0.0726  & 0.1139 & 0.1175 & 0.0904 & \textbf{0.1706} \\
NDCG@5  & 0.0470 & 0.0398 & 0.0526 & 0.0418 & 0.0793 & 0.0833 &  0.0645& \textbf{0.1241} \\
NDCG@10 & 0.0526 & 0.0456 & 0.0581 & 0.0476  & 0.0833 & 0.0880  &0.0681 &\textbf{0.1292} \\ 
\bottomrule
\end{tabular}
}
\end{table*}

\subsection{Evaluation on Trajectory Simulation Task}

\paratitle{Main Results.}
{Table~\ref{tab:tmall_results} reports the performance of our proposed simulator models (Section~\ref{sec:simulation-agent}) against various competing models on the \textsc{Tmall} dataset, from which the following observations can be found:}

First, traditional sequential models (\eg SASRec, GRU4Rec, and BERT4Rec) exhibit relatively limited performance. 
Sole reliance on uniform behavior modeling limits the capacity of these models to capture the underlying user motivations that drive diverse user behaviors in complex e-commerce scenarios.
Second, multi-behavior models (\eg PBAT, MBHT, and MBSTR) consistently outperform traditional baselines by integrating heterogeneous interaction modeling. 
Among them, MBSTR achieves the strongest performance, demonstrating the importance of modeling correlations between different behavior types for intent identification.
Regarding reasoning-based models, ReaRec surpasses SASRec across multiple metrics by introducing a latent reasoning stage to analyze the user's state before the final prediction. 
However, this method still follows the discriminative paradigm, which directly maps historical behaviors to prediction results.

In contrast, our proposed simulator model achieves best performance across all evaluation metrics. 
Unlike existing methods, our framework explicitly simulates the user's decision-making trajectory, characterizing the dynamic evolution from ``divergent exploration'' to ``intent convergence''. 
While multi-behavior models capture diverse interaction signals, they do not model the continuous interest exploration process before making a final decision.
By generating and analyzing multi-step user trajectories, our model decodes the complex behavioral logic before the purchase decision. In empirical evaluations, our approach demonstrates a significant performance improvement over the most powerful baseline models. 
This result underscores the critical value of explicitly modeling the ``exploration-to-decision'' process for enhancing recommendation accuracy.

\begin{table}[t]
  \centering
  \caption{Ablation study of the proposed trajectory simulation agent on the \textsc{Tmall} dataset. ``CR'' and ``PR'' denote the constraint reward and process reward, respectively.}
  \label{table:ablation}
  \begin{tabular}{lcccc}
    \toprule
    \textbf{Model} & \textbf{Recall@5} & \textbf{Recall@10} & \textbf{NDCG@5} & \textbf{NDCG@10} \\
    \midrule
    Ours     & \textbf{0.1557} & \textbf{0.1706} & \textbf{0.1241} & \textbf{0.1292} \\
    \midrule
    w/o CR             & 0.1439          & 0.1651          & 0.1142          & 0.1211          \\
    w/o PR             & 0.1454          & 0.1643          & 0.1165          & 0.1227         \\
    w/o RL             & 0.1187          & 0.1376          & 0.0922          & 0.0982          \\
    \bottomrule
  \end{tabular}
\end{table}

\paratitle{Ablation Study.}
To evaluate the contribution of each core component in our user trajectory simulator, we conduct an ablation study by comparing the complete model design against several variants. The results are presented in Table~\ref{table:ablation}.

$\bullet$ \textbf{w/o CR:} This variant removes the constraint reward defined in Eq.~\eqref{equ:cr}. The constraint reward ensures logical coherence by penalizing trajectory length deviations and structural anomalies. The significant performance degradation indicates that the structural integrity of generated trajectories is crucial for downstream prediction. This finding establishes that high-quality trajectory modeling is a prerequisite for capturing complex user decision-making processes.
    
$\bullet$ \textbf{w/o PR:} This variant excludes the process reward in Eq.~\eqref{equ:pr} and only leverages sparse outcome reward. 
The drop in performance indicates that outcome reward alone is insufficient to discriminate the quality of divergent exploration trajectories. 
By optimizing collaborative consistency in the latent space rather than enforcing rigid token-level matching, the process reward provides essential intermediate guidance while enabling the model to generalize to diverse and reasonable exploration behaviors.

$\bullet$ \textbf{w/o RL:} This variant bypasses the reinforcement learning stage and only adopts SL for model optimization. The result reveals that SL struggles to enhance model generalization and achieve diverse exploration.
The RL stage significantly enhances the stability and logical consistency of trajectory generation through multi-trajectory sampling and policy optimization.

Overall, the ablation results demonstrate that each component plays a critical and complementary role. CR ensures structural coherence, PR guides semantic intent convergence, and the RL stage provides essential robustness for long-horizon generation. Together, these components enable our model to effectively simulate "exploration-to-decision" trajectories, leading to superior recommendation performance.

\begin{wrapfigure}{r}{0.45\textwidth}  
  \centering  
  \includegraphics[width=0.95\linewidth]{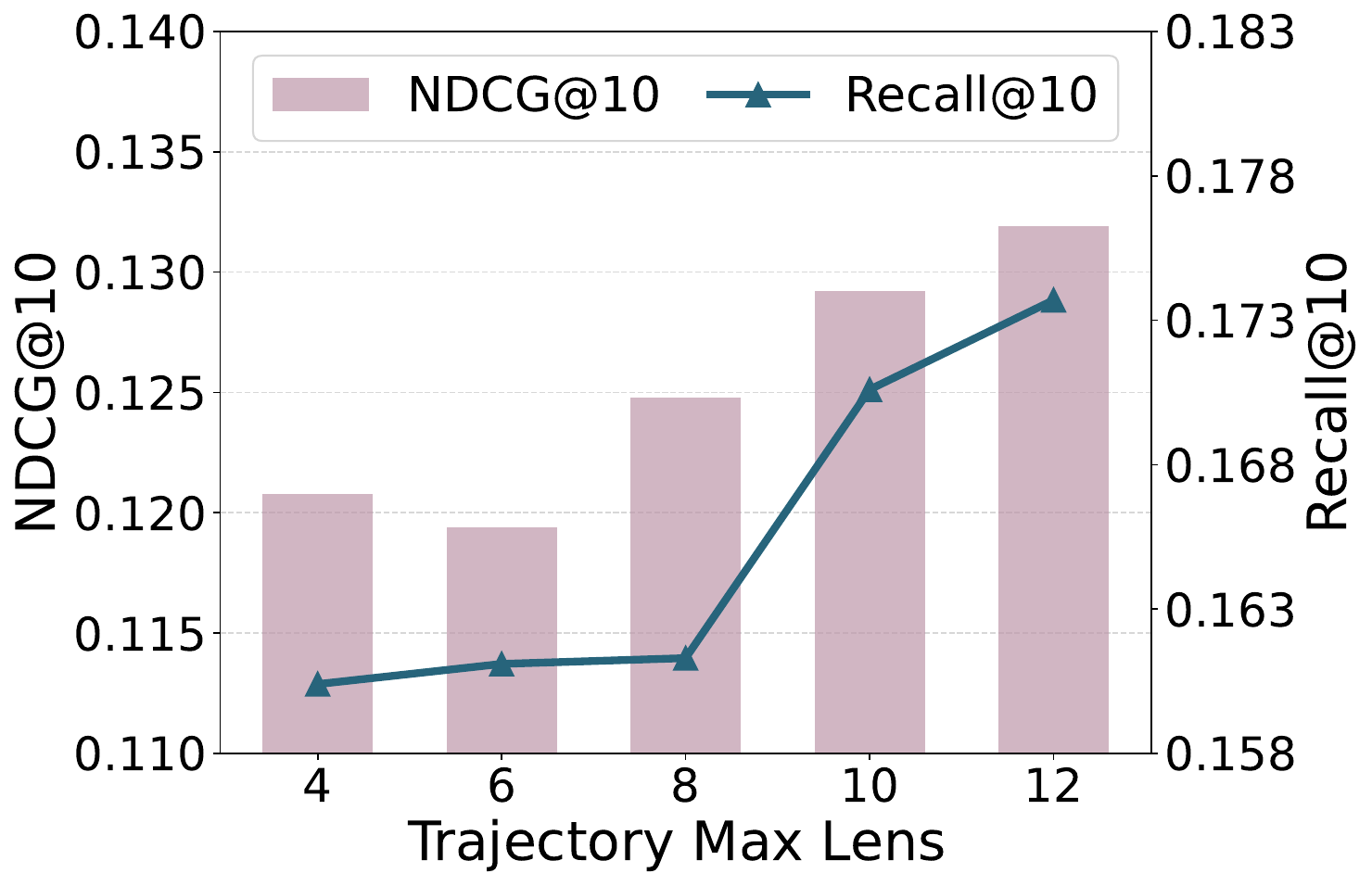}  
  \caption{Performance \wrt maximum trajectory lengths on the \textsc{Tmall} dataset.}  
  \label{fig:path}          
\end{wrapfigure}

\paratitle{Impact of Trajectory Simulation Depth.}
We further investigate how the maximum generation length affects model performance. 
As illustrated in Fig.~\ref{fig:path}, there is a clear positive correlation between trajectory length and recommendation performance.
The length of the simulated trajectory determines the depth of the reasoning process. 

The longer trajectory allows the model to simulate a complete decision-making process by a progressive transition from broad category-level exploration to fine-grained interest convergence.
This progressive accumulation of contextual information enables more precise intent inference. 
Conversely, reducing the maximum length severely curtails the available reasoning context, forcing the model to make premature decisions with insufficient information. 
This results in an insufficient characterization of the user's true intentions, leading to performance decline.
These findings emphasize that explicit exploration trajectory simulation can enable models to progressively capture complex user preferences, thereby enhancing their capability for final decision prediction.

\paratitle{Hyperparameter Analysis.}
\begin{figure}[t]
  \centering
  \includegraphics[width=0.95\linewidth]{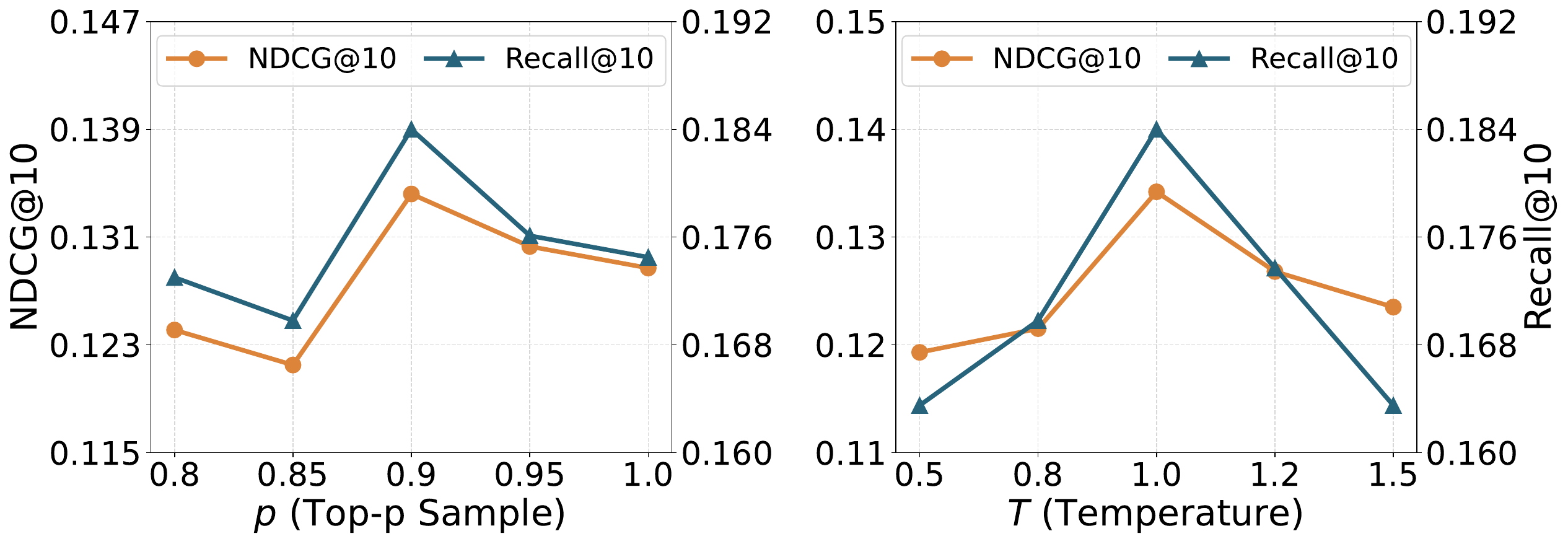}
  \caption{Performance comparison \wrt two sampling parameters on the \textsc{Tmall} dataset: threshold $p$ and temperature $\tau$.}
   \label{fig:hyper}
\end{figure}
To investigate how trajectory diversity affects recommendation performance, we evaluate the impact of two key sampling parameters: the token sampling threshold $p$ and the temperature $\tau$. The results are presented in Figure~\ref{fig:hyper}.

$\bullet$ \textbf{Impact of Top-$p$:} A smaller $p$ (\eg $p=0.8$) overly restricts the sampling space, resulting in a conservative generation process that fails to produce diverse and informative candidate trajectories. 
As $p$ increases, the model gains greater freedom to explore, leading to substantial improvements in recall and ranking quality. 
However, an excessively large $p$ introduces low-probability noise tokens, which undermines the semantic consistency and stability of the simulated trajectory and ultimately degrades performance.

$\bullet$  \textbf{Impact of Temperature $\tau$:} The model performance exhibits a similar trend as $\tau$ changes, peaking at $T=1.0$. Low values result in overly sharp probability distributions, causing the model to favor high-probability yet potentially suboptimal patterns. 
Moderately increasing $\tau$ helps the model escape local optima and generate more informative exploration trajectories. Conversely, excessively high $\tau$ makes the distribution too smooth, and the resulting randomness adversely affects recommendation accuracy.

\subsection{Evaluation on Report Generation Task}

\begin{table}[t]
\centering
\caption{Performance comparison on the \textsc{Tmall} dataset about the report generation task. The best and second-best results are marked in bold and underlined, respectively.}
\label{table:report}
\resizebox{\textwidth}{!}{
\begin{tabular}{llccccccc}
\toprule
\textbf{Evaluator} & \textbf{Method} & \textbf{Accuracy} & \textbf{Coverage} & \textbf{Informativeness} & \textbf{Clarity} & \textbf{Consistency} & \textbf{Novelty} & \textbf{Avg.} \\
\midrule
\multirow{8}{*}{Simulated User} 
& DeepSeek-V3.2 & 4.46 & 3.44 & 3.67 & 4.50 & 3.88 & 2.86 & 3.80 \\
& GLM-5 & 4.38 & 3.30 & 3.47 & 4.24 & 3.70 & 2.73 & 3.64  \\
& Qwen3-Max & 4.05 & 3.24 & 3.66 & 4.48 & 3.78 & 2.91 & 3.68 \\
& GPT-5.2 & 4.35 & 3.40 & \underline{3.77} & \underline{4.67} & \underline{3.91} & 2.89 & 3.83  \\
\cmidrule{2-9}
& ReAct & 4.45 & 3.35 & 3.62 & 4.41 & 3.82 & 2.92 & 3.76  \\
& Plan-and-Solve & \underline{4.48} & \underline{3.52} & 3.76 & 4.55 & 3.90 & \underline{3.07} & \underline{3.88}  \\
\cmidrule{2-9}
& \method & \textbf{4.60} & \textbf{3.62} & \textbf{3.80} & \textbf{4.80} & \textbf{3.94} & \textbf{4.09} & \textbf{4.14}  \\
\midrule
\multirow{8}{*}{Real User}
& DeepSeek-V3.2 & \underline{4.25} & 3.42 & 3.79 & 4.35 & \underline{4.04} & 2.95 & 3.80  \\
& GLM-5 & 4.13 & 3.19 & 3.57 & 4.10 & 3.85 & 2.76 & 3.60  \\
& Qwen3-Max & 3.76 & 3.29 & 3.77 & 4.35 & 3.94 & 2.94 & 3.67  \\
& GPT-5.2 & 4.07 & 3.41 & \textbf{3.83} & \underline{4.54} & \textbf{4.05} & 2.97 & 3.81  \\
\cmidrule{2-9}
& ReAct & 4.17 & 3.26 & 3.65 & 4.20 & 3.98 & 2.95 & 3.70  \\
& Plan-and-Solve & \underline{4.25} & \underline{3.45} & \textbf{3.83} & 4.41 & 4.00 & \underline{3.13} & \underline{3.84}  \\
\cmidrule{2-9}
& \method & \textbf{4.36} & \textbf{3.60} & \underline{3.82} & \textbf{4.69} & 4.00 & \textbf{4.06} & \textbf{4.09}  \\
\bottomrule
\end{tabular}
}
\end{table}

\paratitle{Main Results.}
Table~\ref{table:report} presents the results of automatic and human evaluation of our proposed report generation agent (Section~\ref{sec:report-agent}).
By categorizing the baselines into direct reasoning models (\eg \texttt{GPT-5.2}, \texttt{DeepSeek-V3.2}) and agent-based frameworks (\ie ReAct and Plan-and-Solve), the evaluation reveals several key insights.
First, agent-based methods generally outperform direct reasoning baselines across most metrics. This validates the premise that synthesizing a comprehensive recommendation report is a complex cognitive task that inherently benefits from iterative, agentic workflows rather than single-step zero-shot prompting.
Second, despite the improvements brought by general-purpose agents, our proposed \method consistently achieves the best overall performance.

In particular, \method establishes a substantial margin in the dimension of Novelty.
This superiority can be formally attributed to our proposed multi-aspect interest decomposition strategy, which explicitly deconstructs user interest into multiple granular subsets of attributes.
By disentangling general preferences from niche and emerging interests, this formulation allows the agent to capture fine-grained user intentions that are often overlooked by general task-planning baselines.
Consequently, \method can generate more diverse and unexpected recommendations while maintaining rigorous factual relevance.

{The Cohen's Kappa between simulated users and real users is {0.7064}, indicating good agreement.
To facilitate large-scale evaluations within a limited budget, unless otherwise specified, we use simulated users for the scoring of each metric in the following parts.
}

\begin{figure*}[t]  
    \centering  
    \begin{subfigure}[b]{.49\textwidth}  
        \centering
        \includegraphics[width=\linewidth]{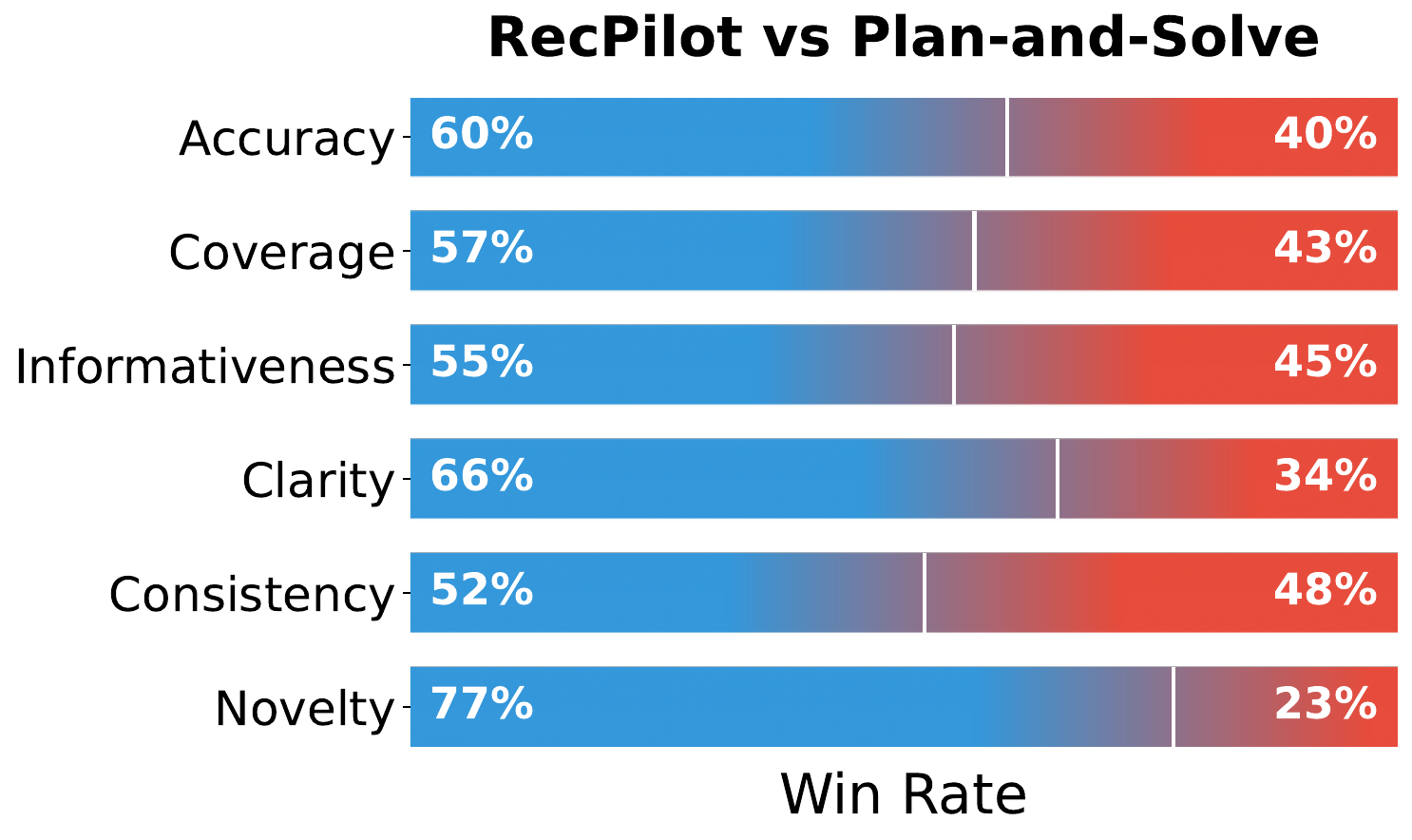}  
        \caption{Pairwise comparison between Plan-and-Solve and our approach.}  
        \label{fig:pair}  
    \end{subfigure}
    \hfill  
    \begin{subfigure}[b]{.49\textwidth}
        \centering
        \includegraphics[width=0.95\linewidth]{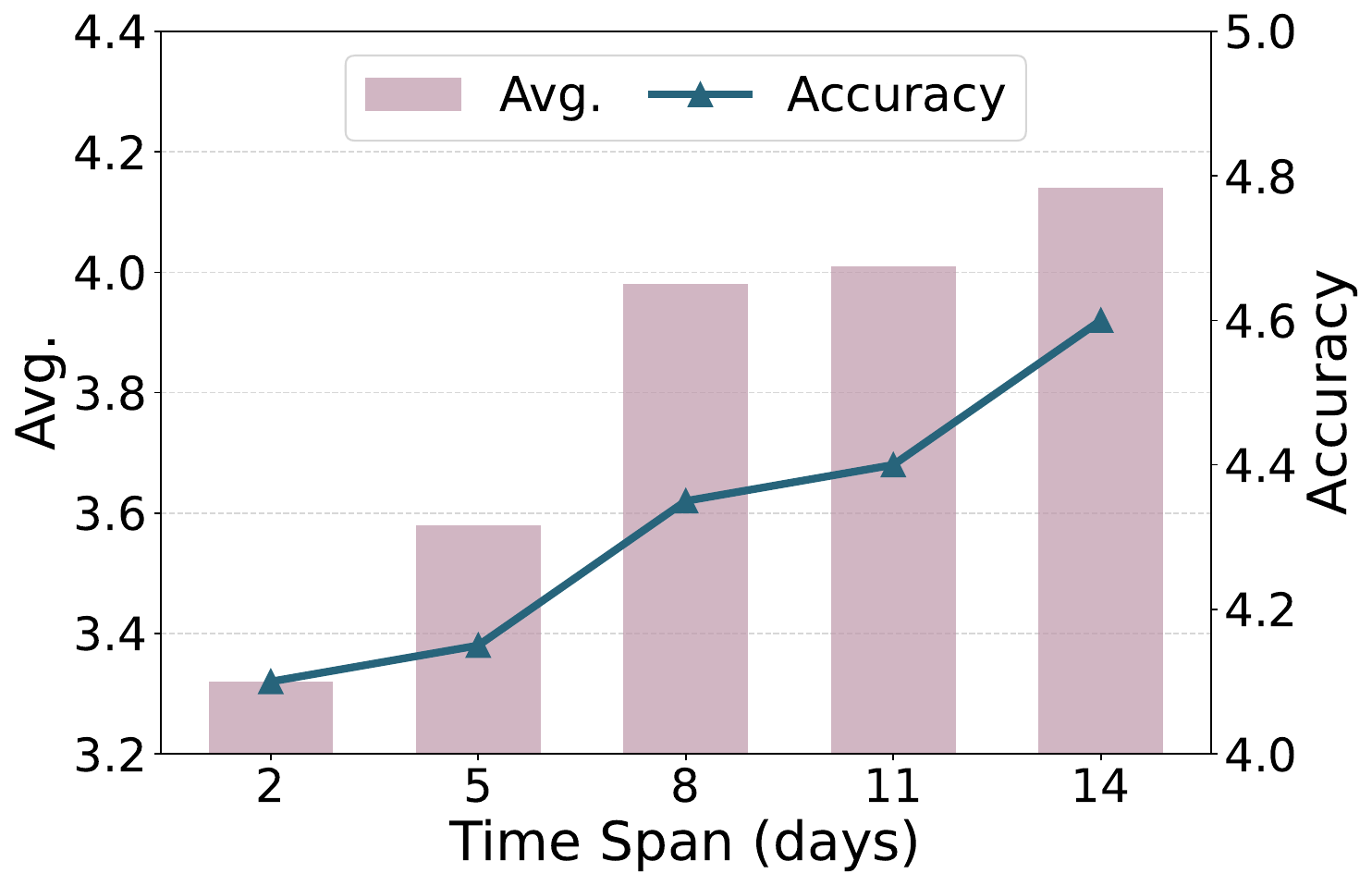}
        \caption{The average scores using interaction records over different time spans for training.}  
        \label{fig:evolution}  
    \end{subfigure}
    
    \caption{Detailed analysis on the \textsc{Tmall} dataset about the report generation task.}
\end{figure*}

\paratitle{Pairwise Comparison.}
To further validate the superiority of our method, we conduct a head-to-head comparison against the strongest agent-based baseline, Plan-and-Solve.
For each test case, we utilize both methods to generate reports and instruct simulated users to select the ``better'' one, subsequently calculating the average win rate.
The comparison results are shown in Figure~\ref{fig:pair}. 
As illustrated in the horizontal bar chart, our approach consistently achieves a higher win rate across all six evaluation metrics. 
Notably, the dominant 77\% win rate in Novelty provides strong quantitative backing for our earlier observation regarding the efficacy of multi-aspect interest decomposition. 
Beyond Novelty, our method demonstrates a decisive 66\% win rate in Clarity. This advantage stems from our aspect-aware parallel ranking design; instead of generating a monolithic summary, \method presents distinct, aspect-specific ranked lists, enabling unambiguous item comparisons that allow users to make direct decisions.
Furthermore, the solid wins in Accuracy (60\%) and Coverage (57\%) validate that our rubric-experience dual-channel mechanism successfully grounds the generated content in both quantifiable attributes and implicit historical contexts, offering more precise and personalized decision support than general-purpose reasoning pathways.






\begin{table}[t]
  \centering
  \caption{Ablation study of the proposed report generation agent on the \textsc{Tmall} dataset.}
  \label{table:ablation_report}
  \resizebox{\textwidth}{!}{
  \begin{tabular}{lccccccc}
    \toprule
    \textbf{Method} & \textbf{Accuracy} & \textbf{Coverage} & \textbf{Informativeness} & \textbf{Clarity} & \textbf{Consistency} & \textbf{Novelty} & \textbf{Avg.} \\
    \midrule
    \method & \textbf{4.60} & \textbf{3.62} & \textbf{3.80} & \textbf{4.80} & \textbf{3.94} & \textbf{4.09} & \textbf{4.14} \\
    \midrule
    w/o Rubrics  & 4.53 & 3.57 & 3.77 & 4.76 & 3.90 & 4.08 & 4.10           \\
    w/o Experience  & 4.49 & 3.55 & 3.76 & 4.74 & 3.87 & 4.06 & 4.08          \\
    w/o Interest Decomposition   & 4.46 & 3.56 & 3.79 & 4.79 & 3.87 & 4.05 & 4.09           \\
    \bottomrule
  \end{tabular}
  }
\end{table}

\paratitle{Ablation Study.}
{To assess the contribution of each core component in our report generation agent, we conduct an ablation study by comparing the complete model design against several variants.
The results are presented in Table~\ref{table:ablation_report}.
}

$\bullet$ \textbf{w/o Rubrics:} This variant removes the structured rubrics and only utilizes experience for ranking. The consistent decline across all metrics indicates that the absence of unified attribute dimensions undermines the logical architecture of the generated content. Structured rubrics provide a quantifiable measurement for comparison, thereby significantly enhancing the organization and consistency of the report.

$\bullet$ \textbf{w/o Experience:} This variant excludes the textual experience memory and only utilizes rubrics for ranking. The drop in performance suggests that relying solely on explicit attribute scores is insufficient to capture context-dependent implicit user preferences. By providing semantically rich nuances from historical interactions, the experience module serves as a flexible supplement to structured scoring, which is crucial for improving the overall informativeness and robustness of the report.

$\bullet$ \textbf{w/o Interest Decomposition:} This variant removes the multi-aspect interest decomposition strategy, ceasing to decouple user interests into multiple fine-grained aspects based on rubric subsets. The significant drop in Accuracy indicates that deconstructing macro-level user preferences into specific interest slices is pivotal for capturing complex user intents.

\paratitle{The Effect of Self-Evolution.}
In this part, we evaluate the self-evolving ability of our report generation agent, which performs continuous optimization of rubrics and accumulation of experience.  
Specifically, we leverage interaction records over different time spans (\ie 2, 5, 8, 11, 14 days) for optimization and then compare their performance on the test set.
Results are presented in Figure~\ref{fig:evolution}.
As we can see, the performance of our agent exhibits a consistent upward trend as the time span increases, providing direct evidence for the effectiveness of the proposed self-evolution mechanism.
Thanks to the design of rubrics, the self-evolution module can continuously optimize the their weights using user behavior data. 
The more data there is, the more stable these weight values become, thereby enabling more precise long-term preference modeling.
Especially, in the early stage (\ie from 2 days to 8 days), the agent achieves rapid performance improvement within a few rounds.
This can be attributed to the experience memory module, which can extract meaningful textual preference signals from limited feedback, thereby enabling the agent to quickly adapt to similar scenarios.

\begin{figure*}[t]
    \centering
    \begin{subfigure}[b]{0.33\textwidth}
        \centering
        \includegraphics[width=\linewidth]{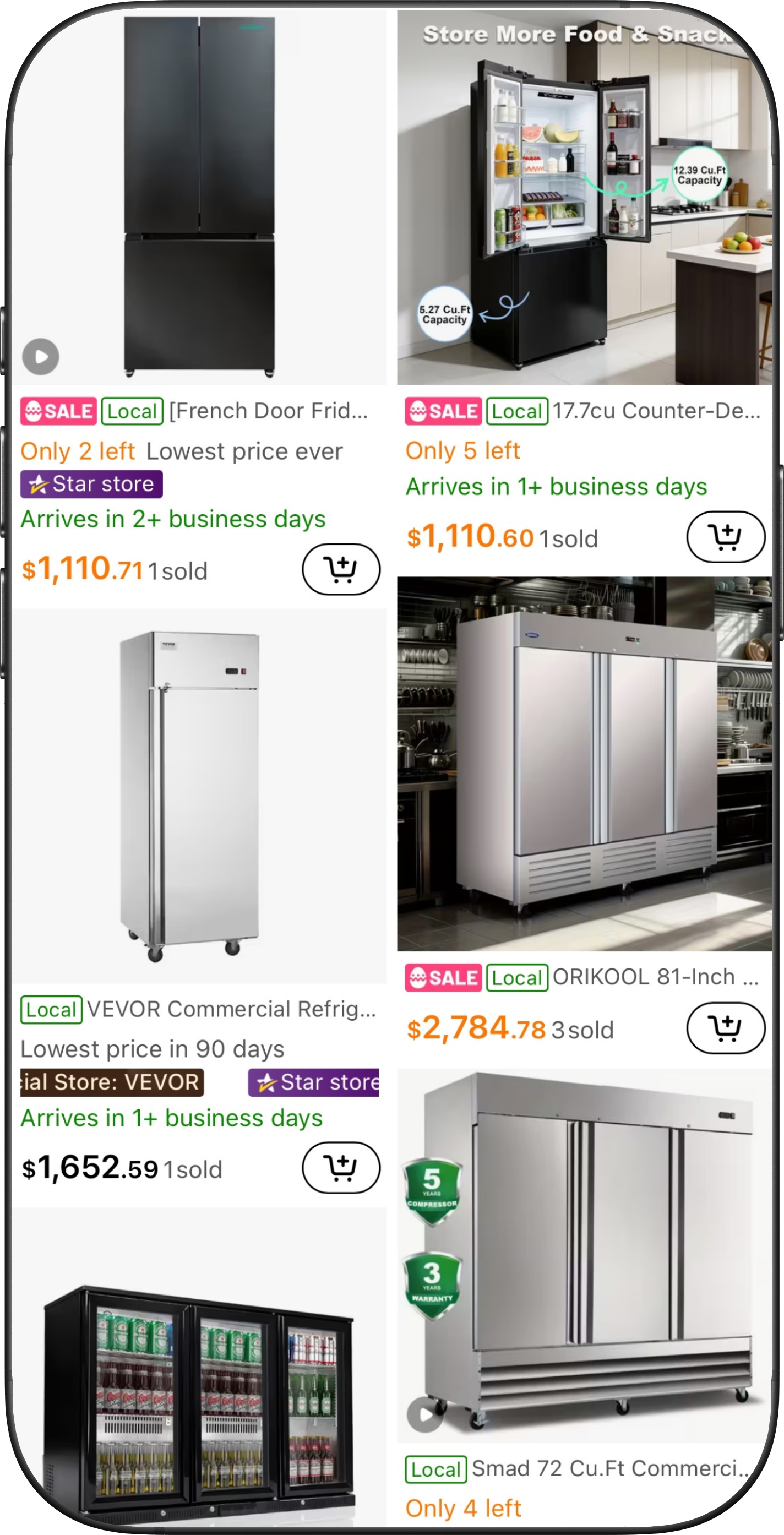}
        \caption{Traditional item lists.}
        \label{fig:case_baseline}
    \end{subfigure}
    \hfill
    \begin{subfigure}[b]{0.63\textwidth}
        \centering
        \includegraphics[width=\linewidth]{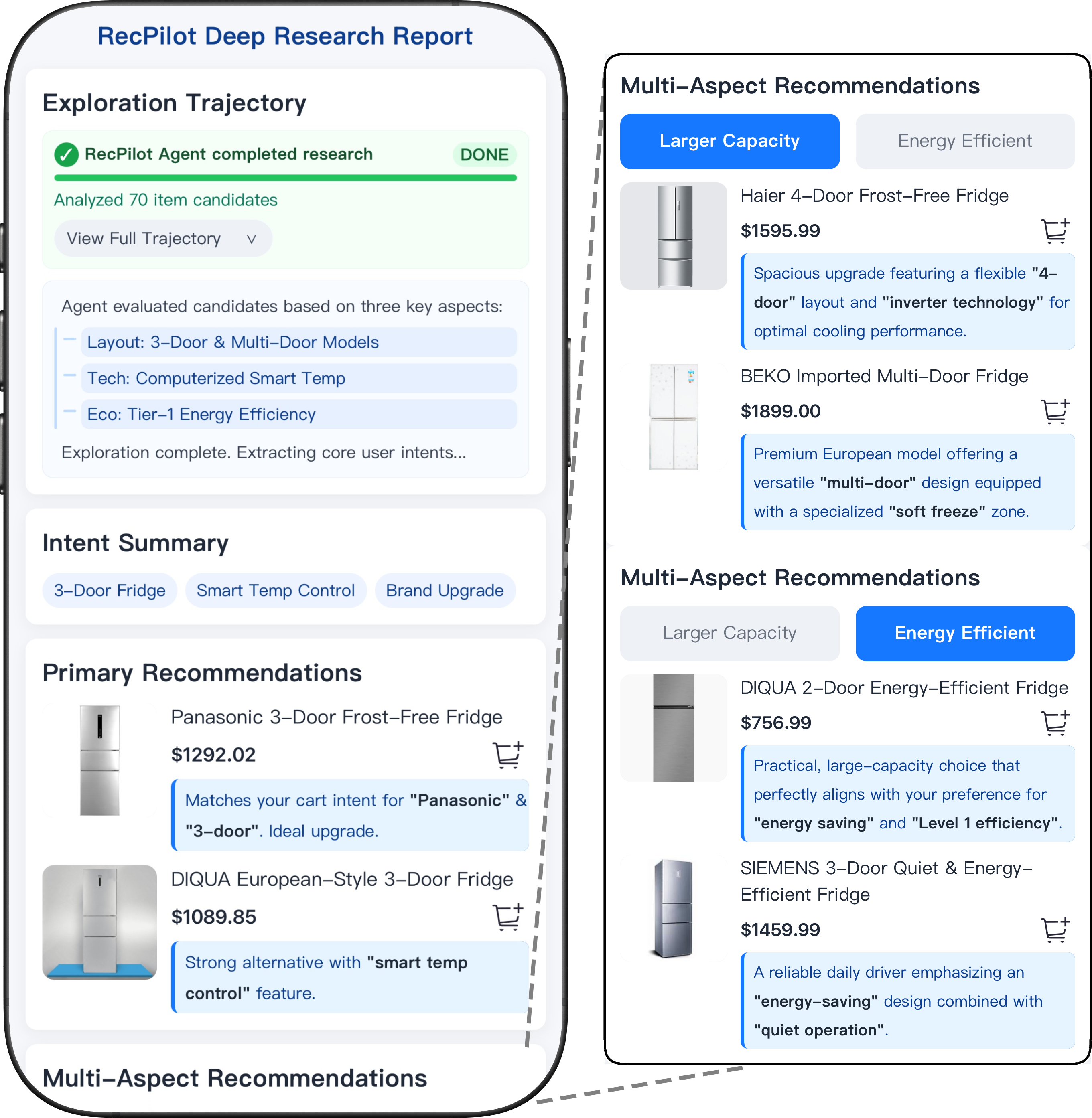} 
        \caption{Report generated by our \method.}
        \label{fig:case_agent}
    \end{subfigure}
    \caption{Sample interaction interfaces in existing recommender systems and our proposed \method.}
    \label{fig:case_study}
\end{figure*}

\paratitle{Case Study.}
To intuitively illustrate how our proposed \method transforms the recommendation experience from passive item exposure to proactive, synthesized reporting, we provide a concrete example in Figure~\ref{fig:case_study}.
In this scenario, a user intends to purchase a refrigerator.
As illustrated in Figure~\ref{fig:case_baseline}, existing recommender systems usually display potentially interesting refrigerators in the form of an item list.
Such a list only contains limited information, including product images, titles, and prices.
Consequently, users are required to click on each item individually to inspect detailed specifications, which is undoubtedly a labor-intensive process that adversely impacts the user experience.

Figure~\ref{fig:case_agent} presents the report generated by our \method for the same scenario. Instead of conventional item lists, the agent autonomously synthesizes its findings into a coherent and interpretable report. It first provides an ``Exploration Trajectory'' section that displays the item exploration process, which aims to enhance the credibility of the report. Next, the ``Intent Summary'' section summarizes the core attributes (\eg 3-door fridge) in which the user may be interested. Subsequently, through multi-aspect interest decomposition, the agent delivers tailored recommendations. It provides ``Primary Recommendations'' to facilitate quick decision-making, as well as ``Multi-Aspect Recommendations'' that explicitly contrast different decision dimensions—such as upgrading to a larger capacity versus prioritizing energy efficiency. This structured presentation significantly alleviates information overload, successfully transforming recommendation from passive, user-driven filtering into proactive, agent-driven services.
\section{Related Work}


\subsection{LLM-Based Recommendation}

Large language models (LLMs) have been widely applied to recommender systems, benefiting from their advanced semantic understanding and reasoning capabilities~\cite{llm_survey,llmrec_survey1,tallrec}.  
Initial investigations primarily adopted zero-shot or few-shot paradigms, reformulating tasks such as sequential recommendation~\cite{llmrank,DBLP:conf/recsys/DaiSZYSXS0X23}, user data augmentation~\cite{kar,DBLP:journals/corr/abs-2401-13870}, and conversational recommendation~\cite{chatrec} into textual prompts.  
However, the divergence between general language modeling and domain-specific recommendation tasks often remains challenging to bridge through prompt engineering.   
Consequently, subsequent studies have focused on better adapting LLMs to recommendation scenarios.  
This includes instruction tuning on recommendation data to enhance performance~\cite{p5,lc-rec,instructrec}, integrating preference optimization to better align with user preferences~\cite{s-dpo,sprec}, and employing reinforcement learning to improve deep reasoning within recommendation contexts~\cite{deeprec,latentr3}.  
Furthermore, some advancements have also attempted to develop LLM-based recommendation agents~\cite{agentcf,InteRecAgent}, utilizing techniques such as multi-agent frameworks~\cite{InstructAgent,macf}, memory modeling~\cite{agentcf++,mrrec} and reinforcement learning~\cite{mgfrec,starec}.

\subsection{Deep Research}
Recently, large language models have evolved into powerful problem solvers, but complex open-ended tasks often exceed the capabilities of single prompts or standard retrieval-augmented generation (RAG)~\cite{rag_survey}.
This limitation has led to the emergence of the deep research (DR) paradigm, which frames LLMs as autonomous research agents through end-to-end workflows~\cite{dr_survey,openai_dr}.
A typical DR system integrates four core components:
Query planning decomposes complex queries into executable sub-tasks via parallel, sequential, or tree-based strategies~\cite{DBLP:conf/iclr/ZhouSHWS0SCBLC23,search-r1,r1-searcher}.
Adaptive information acquisition involves invoking retrieval tools, adaptive search time planning, and information filtering~\cite{search-o1,DBLP:journals/corr/abs-2505-20128}.
Dynamic memory management governs the memory lifecycle via consolidation, indexing, updating, and forgetting mechanisms~\cite{memorybank,a-mem,mem0}.
Answer generation synthesizes upstream information, resolves conflicting evidence, and logical reasoning~\cite{react}.
Recent advancements further optimize DR through techniques such as agentic prompting~\cite{gemini_dr,openai_dr}, supervised fine-tuning~\cite{webthinker,evolvesearch}, and end-to-end reinforcement learning~\cite{r1-searcher,DeepResearcher}.

\section{Conclusion}
In this work, we  consider a long-standing limitation in recommender system design:  the final output returnes a list of items for the user to evaluate, which  places an undue cognitive burden on users, forcing them to act as their own information synthesizers. To address this, we proposed a fundamental shift in interaction—from passive item exposure to proactive, synthesized reporting. We instantiated this vision with \method, a multi-agent framework that autonomously explores the item space and generates interpretable, user-centric reports to directly support decision-making. Our experiments validate that this approach not only effectively models user preferences but also produces coherent and persuasive outputs that demonstrably reduce user effort.

When considering the future deployment of our approach, we recognize two key aspects that warrant careful attention. First, not all product categories are equally suited to this new interaction paradigm; we anticipate the greatest value for high-priced items, where users typically invest considerable time in making final decisions. Second, some users may prefer faster interactions and could perceive the report generation process as more time-consuming compared to traditional list-based recommendations. In light of these considerations, we believe a dual-mode system—integrating both fast, traditional recommendations and slower, reasoning-based report generation—would offer significant value. In such a design, users would have the option to activate the more deliberative mode when they seek deeper decision support.


Looking ahead, this work opens several promising directions. The emerging deep research paradigm points toward a potential future in which recommender systems evolve from simple filters into intelligent assistants—capable of reasoning about user needs and justifying their recommendations. Key challenges remain, including improving the efficiency of trajectory simulation for large-scale item pools, ensuring the factual consistency of generated reports, and developing robust evaluation metrics that go beyond traditional accuracy measures to capture user cognitive load and satisfaction. We believe this work lays the foundation for a new generation of recommender systems that are not only more effective but also more helpful—ultimately freeing users from the burden of information seeking.




\bibliographystyle{unsrt}
\bibliography{ref}

\appendix
\section{Appendix}

\subsection{Details of Human Evaluation}
\label{app:human-evaluation}

To mitigate the inherent subjectivity of human evaluation and ensure the reproducibility of the results, we developed a strict human evaluation protocol.
Evaluators were required to assess the generated reports based on explicit behavioral anchors rather than holistic impressions.
The detailed guidelines are provided below.

\subsubsection{Task Setup}

We designed a blind-test evaluation task to assess whether the generated reports can effectively assist in decision-making and reduce the burden of exploration.
For each test case, human evaluators were provided with historical interactions, the simulated exploration trajectory, and the set of candidate items to serve as the evaluation context.
The reports generated by different baseline models and the proposed approach were strictly anonymized and presented in a randomized order.
Evaluators were then required to score each report independently across six predefined dimensions, using the provided evidence and the ground truth purchased item as the absolute anchor for accuracy.

\subsubsection{Evaluator Instructions}

Evaluators were instructed to strictly adhere to the following core rules to minimize subjective bias:

$\bullet$ Evidence-Based Grading: Evaluators must rely solely on the provided historical trajectories and candidate sets as evidence. Any specific item attributes or behavioral details mentioned in the report that cannot be traced back to the input must be penalized as hallucinated content.

$\bullet$ Ground Truth Anchoring: The ``best choice'' is defined strictly as the actual purchased item. Recommending items outside the candidate set, or prematurely referencing the removed ground truth purchase action, must result in severe penalties.

$\bullet$ Tolerance for Simulated Tone: Since the target session is generated via trajectory simulation, reports should not be penalized for using a speculative or simulated tone. However, asserting the simulated trajectory as a past factual event should be appropriately penalized.

$\bullet$ Form versus Function: The evaluation must focus entirely on whether the report genuinely assists in decision-making. Evaluators must not be swayed by specific terminology, stylistic templates, or mere grammatical fluency if the underlying logic is flawed.

\subsubsection{Scoring Rubric}

To ensure high inter-annotator agreement, we developed a fine-grained 5-point scale with explicit behavioral anchors for each evaluation metric. Evaluators were required to match the quality of the report to the specific descriptions outlined in Table~\ref{tab:human_eval_rubric}.

\begin{table*}[h]
\centering
\caption{The 5-Point Scoring Rubric for Human Evaluation with Explicit Behavioral Anchors.}
\label{tab:human_eval_rubric}
\begin{tabular}{m{0.15\linewidth}  m{0.25\linewidth}  m{0.25\linewidth}  m{0.25\linewidth}}
\toprule
\textbf{Metric} & \textbf{Score 1 (Poor)} & \textbf{Score 3 (Acceptable)} & \textbf{Score 5 (Excellent)} \\
\midrule
\textbf{Accuracy} & Completely misses the label or highly similar items in both primary and secondary recommendations. & Primary recommendation misses the label, but the label or a highly similar item is included in the secondary recommendations. & Primary recommendation accurately hits the label or a perfectly aligned substitute. \\
\midrule
\textbf{Coverage} & Lacks key decision factors or relies entirely on fabricated features without evidence. & Mentions some relevant clues, but lacks focus or the evidence chain is weak. & Explicitly highlights 2-3 decisive clues and firmly binds them to candidate comparisons. \\
\midrule
\textbf{Informativeness} & Filled with meaningless filler words, generic templates, or severe hallucinations. & Contains useful information but is mixed with generic redundancy and lacks deep insight. & High information density; every claim is strictly backed by the input evidence. \\
\midrule
\textbf{Clarity} & Highly ambiguous, contradictory, or fails to draw a clear conclusion for the user. & Provides a conclusion, but comparisons or conditions for alternative choices are vague. & Primary recommendation is explicit, and conditions for choosing alternatives are distinct. \\
\midrule
\textbf{Consistency} & Severe logical breaks, or directly contradicts the historical trajectory evidence of the user. & Generally aligned with inputs, but contains minor logical jumps or slightly forced deductions. & Perfect fidelity; all deductions and causal links strictly follow the trajectory evidence. \\
\midrule
\textbf{Novelty} & Lacks any forward-looking insights, or bases ``insights'' entirely on hallucinated details. & Offers some insights or warnings, but they are too generic or not critical to the decision. & Proactively points out high-value, evidence-backed risks or opportunities. \\
\bottomrule
\end{tabular}
\end{table*}

\end{document}